**When Screening Current Controls Ferroelectric Switching: From Field-Limited to Current-Limited Regimes**

Denis Alikin[1]*, Anastasya Meng[2], Michael Kosobokov[2], Semen Melnikov[2], Violetta Safina[2]
[1]CICECO Aveiro Institute of Materials & Physics Department, University of Aveiro, Aveiro, Portugal, denis.alikin@ua.pt
[2]School of Natural Sciences and Mathematics, Ural Federal University, Ekaterinburg, Russia

**Abstract**
Tip-induced switching in ferroelectrics is commonly framed as a field-driven process controlled by domain-wall kinetics. Here we argue for a more universal viewpoint: under a scanning probe, polarization reversal is fundamentally constrained by how fast screening charge can be supplied and redistributed. By combining local switching experiments with finite-element simulations, we identify a screening-current-limited mechanism in which charge injection and its space-charge-limited relaxation set the pace of switching and the scaling of domain growth. This framework naturally explains regime changes and the apparent breakdown of "intrinsic" domain-wall laws often inferred in the nanoscale piezoelectric hysteresis measurements. Beyond a reinterpretation of tip switching, the results position screening currents as a hidden control parameter of ferroelectric reversal across materials, turning current regulation into a practical handle for deterministic nanoscale domain engineering.

**Introduction**

Reversal of spontaneous polarization is a fundamental requirement for a material to be classified as ferroelectric. Understanding polarization switching at the nanoscale is critical for a wide range of applications, especially in the rapidly developing field of two-dimensional (2D) materials[1,2] and ferroelectric membranes[3,4], where conventional electrode-based measurements are often challenging or where local manipulations are desired. It is also essential for controlling complex polarization topologies such as vortex domains[5] and domain walls[6], probing the properties of complex heterophase materials[7]. Piezoresponse Force Microscopy (PFM) is an established technique that enables both identification of the polarization orientation and local polarization switching at the nanoscale under a biased scanning probe microscopy (SPM) tip[8].

Significant efforts have been made to quantify the piezoelectric response in PFM[9–11], and substantial progress has been achieved in recent years, by implementing advanced spectroscopic techniques[12,13] and, more recently, interferometric approach[14]. Despite extensive work since the early 2000[15] still, however, the underlying mechanisms of the local polarization switching under SPM tip remain insufficiently understood. One of the main parameters typically accessible from local switching experiments is the coercive voltage extracted from Switching Spectroscopy PFM (SSPFM) loops, which often differs substantially from values obtained in capacitor geometries. This discrepancy has recently been attributed to the presence of a nanoscale dielectric gap between the surface and the probe[16].

Another common approach in local switching studies with an SPM tip is to interpret the dependence of domain radius formed after local switching on the pulse duration as a time-resolved observation of domain wall motion[17–22]. Many prior studies have focused on fitting these dependencies using models that combine the point-charge approximation for the electric field from the tip and empirical activation[17–19] or linear[20] dependence of the domain wall velocity on the electric field. However, estimated activation field for domain walls motion is often found to be quite different from that measured in capacitor geometry[21,23] while the measurements across very different materials often yield similar final domain radii[24]. Many of the earlier studies also reported a linear dependence of domain radius on the amplitude of the applied voltage pulses[18–20]. Nevertheless, this result still lacks a consistent physical interpretation. In a simplified model, assuming a linear dependence of domain wall velocity on electric field and a point-charge field distribution, one can derive a theoretical $r \sim V^{1/n}$ with $n \approx$ 3-4 dependence for the stationary solution assuming that for long pulses, domain-wall motion nearly ceases (Supporting Information S1). More advanced thermodynamic models also predict nonlinear saturated behaviors. For instance, Molotskii et al. predicted an equilibrium domain size with a $V^{2/3}$ scaling[25]. The model of Morozovska and Eliseev[26] , incorporating Debye screening, predicts an even more saturated dependence. Rodrigues attributed deviations from linearity to transitional dynamics during switching[18], while Tybell et al.[21] and later Soergel et al.[19] justified linear behavior by assuming uniform electric field from the probe and considering only ultra-thin film case, thereby neglecting important electric field non-uniformity at larger scale. More recent work by Podivilov and Sturman that incorporates conductive domain walls[27,28] also does not reproduce linear and superlinear dependencies on the amplitude of the applied voltage pulses. In particular, measurements performed in air or under humid conditions may also involve various electrochemical phenomena and ferroionic coupling effects, further complicating the overall physical picture[29,30].

This study demonstrates that local switching induced by an SPM tip cannot be adequately explained solely by the non-uniform electric field from the tip and domain-wall kinetic laws. Instead, switching is governed by the polarization-screening pathway: charge injection and its subsequent relaxation through a space-charge-limited current (SCLC) mechanism. Accordingly, in SSPFM the apparent coercive voltage is not necessarily an intrinsic switching threshold but can reflect the transport response of the tip-surface junction. Conventional activation-based analyses of switching kinetics and hysteresis loops may therefore be misleading unless the limiting regime is identified. At the same time, the current-limited regime provides a practical route to control domain shape by regulating the injected current, enabling the fabrication of designed nanoscale ferroelectric structures.

**Experimental details**

Experimental study was conducted on (0001) periodically poled single-crystalline congruent lithium niobate (Crystal Tech., Inc., USA), thinned to ~100 μm. PFM visualization, local polarization switching by rectangular pulses and SSPFM measurements were performed using an Ntegra Aura SPM (NT-MDT, Russia) equipped with NSG01 platinum-coated probes (35 nm tip radius, 5 N/m stiffness). 21 kHz, 3 V AC voltage was used for PFM visualization. A NI-6251 DAQ board (National Instruments corp., USA) and a high-voltage amplifier Trek-677B (Trek, Waterloo, WI, USA) were used to generate voltage pulses in the local switching experiments. The voltage pulses were applied to the probe, with the crystal's backside grounded with a conductive substrate. Local polarization switching was performed using rectangular pulses with durations ranging from 0.1 to 10 s and amplitude ranging from 10 V and 200 V. To avoid polarization backswitching following the reduction of the electric field, the tip was retracted prior to voltage drop. To control the injection current, resistors with nominal values ranging from 100 kΩ to 10 GΩ were series-connected to the electrical circuit between the substrate and ground. A surface potential mapping employed closed-loop KPFM with 0.5 V AC at a 100 nm tip-surface distance. To assess the possibility of switching in the low-voltage region, local hysteresis loops were measured using the positive up-negative down (PUND) method within the 'step-mode' SSPFM technique [31]. 1 V, 21 kHz AC voltage was applied to the probe. The DC voltage peaked at 80 V, with pulse durations ranging from 10 ms to 10 s. All measurements took place in a dry nitrogen atmosphere (<1% humidity), maintained by constant nitrogen flow through the microscope chamber. A subset of the experiments was also reproduced under low-vacuum conditions (~$10^{-3}$ Torr). However, the resulting dependencies showed negligible differences compared with measurements performed in air, and therefore these data are not included in the manuscript.

Finite-element simulations were performed in COMSOL Multiphysics using the Electrostatics module. The electric-field distribution was calculated for a probe with a 50 nm tip radius in contact with the surface. The simulations were made for charge mobility $10^{-10}\div10^{-14}$ m$^2$/(V·s), typical range of theoretically calculated (~$10^{-14}$ m$^2$/(V·s))[32] and experimentally measured for Fe-doped lithium niobate (~$10^{-13}\div10^{-12}$ m$^2$/(V·s))[32]. Charge injection was implemented by applying an ohmic current-flow boundary condition over the probe–sample contact area, with the contact resistance set empirically in the range 10 kΩ÷1 MΩ. Two simulation types were performed: with and without polarization switching. Domain growth was defined by the variation in bound charges in the subsurface layer of 5 nm thickness, ignoring domain wall dynamics in the crystal bulk. The charge was allocated throughout the subsurface layer according to a truncated Gaussian distribution. The domain wall was modeled with a sigmoid function to show the polarization change from $-P_s$ to $+P_s$, featuring a 1 nm-thick transition zone as its effective width. Domain wall dynamics and injected charge motion redistribution were solved as a fully coupled time domain problem. A more detailed discussion of the equations used in the simulations is provided in the main text of the paper.

## Results and discussion

### 1. Local polarization switching experiments

We performed local switching experiments in the periodically poled lithium niobate crystals with an average period of 10 µm, polished to a thickness around ~100 µm. We first started with the measurements inside individual domains and voltage pulses have been applied to regions placed far from the position of the domain walls (Figure 1). Application of the voltage pulse inside the pre-existing domains resulted in the formation of the stable domain only when positive voltage has been applied in $z^+$ domains (surface with the positive bound charge). While application of the negative voltage pulses in $z^-$ domains (surface with the negative bound charge) didn't result in the switching of polarization. Such an 'unipolar' switching was observed earlier and explained by the significantly different electrical conductivity of the domain walls formed by pulses of the different polarity and consequently their different stability[33–35].

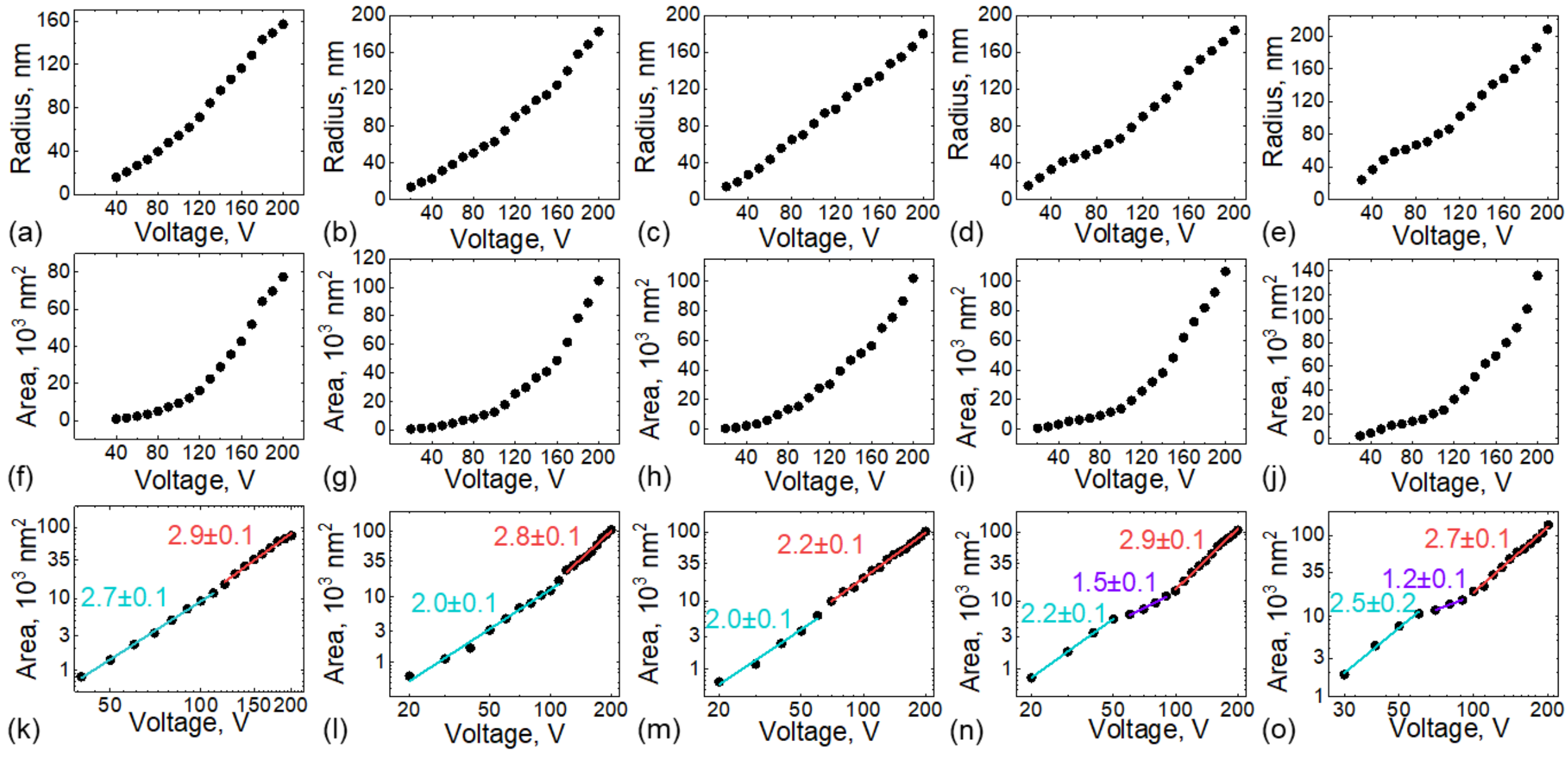

Figure 1. The results of local polarization switching using an SPM tip. Shown are the dependencies on the amplitude of the voltage pulses: (a)-(e) domain radius, (f)-(j) domain area, (k)-(o) domain area plotted in double logarithmic scale; for different pulse durations: (a), (f), (k) 100 ms, (b), (g), (l) 200 ms, (c), (h), (m) 500 ms, (d), (i), (n) 1 s, (e), (j), (o) 10 s.

The voltage for the formation of stable ferroelectric domains, i.e. those possible to be visualized after switching, was found to be in the range of 20 to 40 V, depending on the duration of the voltage pulses. The domain area and radius as function of the amplitude of the applied voltage for various pulse durations are presented in Figure 1. The dependencies varied with pulse duration, reflecting differences in switching dynamics. Generally, the domain radius exhibited a power-law dependence with an exponent in the range of ~1.0–1.4, while the domain area showed a power-law behavior with exponents ~2.0–2.7 (Figure 1k-o), consistent with previously reported results for low amplitudes of the applied voltages. Notably, the power-law exponent increased with voltage amplitude. The respective domain size dependence on pulse duration is shown in Figure 2a,b and Figure S1, as well as discussed further in the main text.

It is likely that below ~20-40 V reversible switching can happen, in which the switched domain relaxes back upon removal of the bias under the action of the depolarization field[33,36,37]. Reversible polarization switching can be evaluated in situ by monitoring piezoresponse during increasing voltage pulses, i.e. SSPFM measurements. In our case, we used the PUND SSPFM approach, which allowed us to exclude the back-switching happening at the voltage ramp-down stage[31] (Supporting Information S3). The complete PUND SSPFM spectroscopy cycle is shown in Figure S2, while Figure 2c presents the piezoresponse as a function of the applied DC voltage. The horizontal lines indicate the threshold above which the piezoresponse exceeds the noise level. The corresponding bias is operationally defined as the apparent threshold voltage for the onset of polarization reversal. The extracted threshold voltages are summarized in Figure 2d. For short voltage pulses, the threshold was about 15 V, whereas longer pulses enabled polarization switching at much lower voltages. This behavior differs markedly from previously reported local switching threshold[19] and indicates that polarization reversal can occur under much lower electric field.

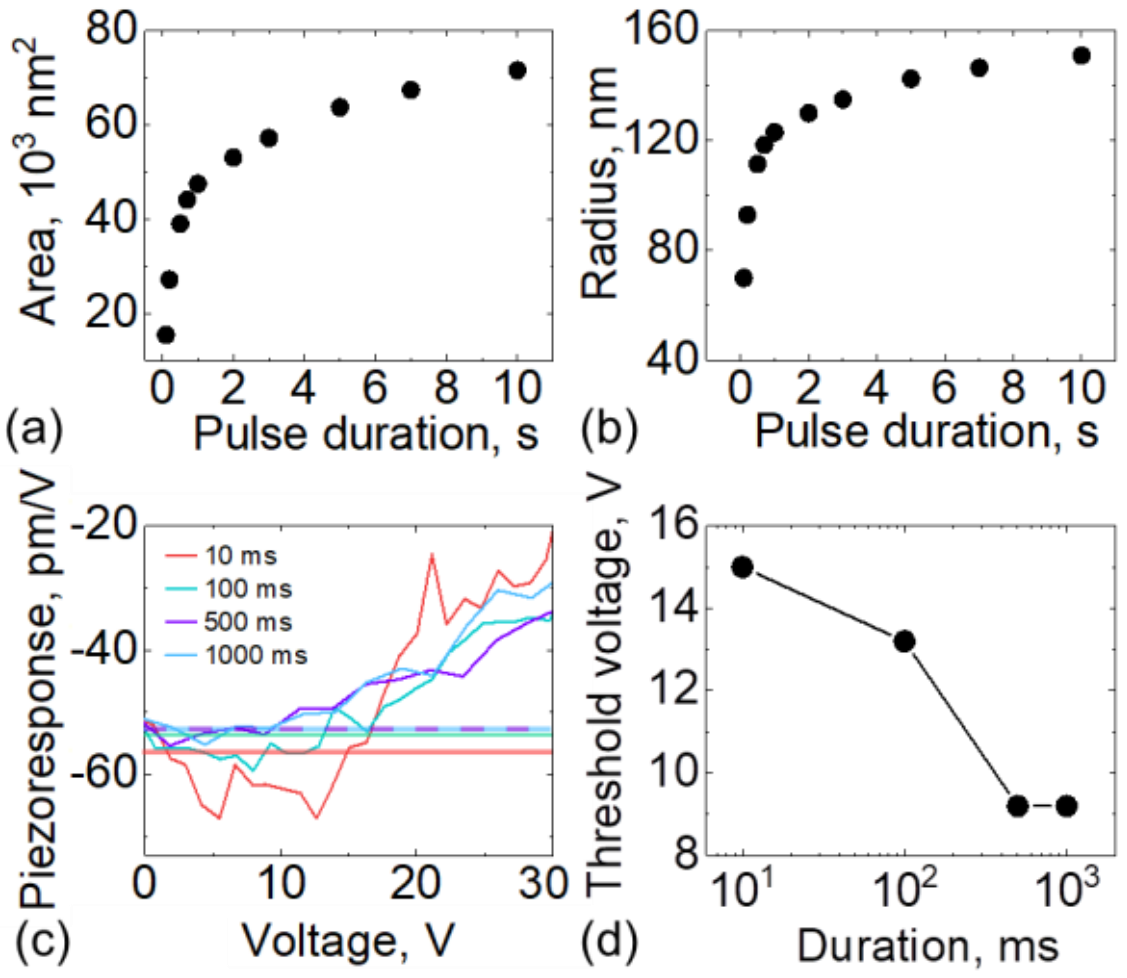

Figure 2. Dependence of domain (a) area and (b) radius on pulse duration for the applied voltage pulse amplitude of 100 V. (c) Piezoresponse dependence on the applied DC voltage. (d) Threshold voltage dependence on the duration of the applied voltage pulses.

The average domain area is directly proportional to the amount of charge required to screen the ferroelectric polarization. In conventional capacitor-based switching, this charge corresponds to the time-integrated displacement (switching) current in the external circuit. However, in the case of SPM tip switching, the situation differs. The switched area exhibited a clear power-law dependence on the applied voltage, with no indication of either linear or exponential scaling (Figure 1k–o). This prevents a simple attribution of the observed behavior to ohmic current flow or to barrier-controlled charge transport. The SPM tip typically does not provide ideal external screening, due to factors such as imperfect contact[38], the presence of a dielectric gap at the surface[16,39], and the extremely high electric field up to $10^{10}$ V/m near the tip. These conditions promote strong charge injection into the ferroelectric surface to locally reduce the field to more stable levels[40]. As a result, the domain area vs. amplitude dependence reflects the voltage dependence of the injected screening charge.

The power-law exponents in the range of 1.5-2.5 are typical of SCLC mechanism, which are characteristic for charge injection from an SPM probe, irrespective of the material[41]. Therefore, we conclude that screening is likely dominated by injected charges from the probe, followed by surface and bulk redistribution, rather than by simple external circuit-mediated charge flow. Space-charge-limited current can be described using the general equation[41]:

$$I_{sclc} = aU^m, \quad (1)$$

where $a$ is a constant depending on the electrical properties of the material and the exponent $m$ characterizes the mechanisms of current flow. The regime with $m = 2$ (Mott-Gurney regime) assumes charge transport in the surface layer with a fully occupied charge traps. The realization of this regime in lithium niobate is highly probable as lithium niobate is strong insulator with a large band gap. In the case of the larger voltages deep traps started to be filled, *m* can increase further greater than *2*, which manifests “modified” Mott-Gurney regime. The latter explains growth of the exponents with the amplitude of the voltage pulses. Finally, regime with $m = 3/2$ (Child’s law) could appear in the case of the massive charge transport into the material bulk with no significant dissipation.

It should also be emphasized that the experiments were performed on lithium niobate under dry nitrogen or low-vacuum conditions. Under these conditions, and at the temperatures used here, substantial ionic redistribution in $LiNbO_3$ is unlikely, while pronounced charge injection and accumulation are clearly observed. This makes electronically dominated transport the most plausible origin of the response, rather than electrochemical kinetics.

**2. Screening charge dynamics**

Following this empirical consideration of the local polarization switching, we further decided to study the relationship between the screening charge dynamics and variation of the domain radius with increasing the amplitude of the voltage pulses. Although measuring very small injection currents with sufficient temporal resolution remains technically challenging, the final charge distribution can be analyzed using a Kelvin probe force microscopy (KPFM) approach. Before performing the local switching experiments, KPFM scan was acquired to assess the initial charge state of the material, which was then used as a reference for background subtraction in the subsequent analysis. Further, we applied positive voltage pulses with the fixed duration and increasing amplitude to the lithium niobate with the different polarization orientation ($z^+$ and $z^-$) and $SiO_2$ thick oxide layer on top of silicon crystal. After voltage pulses application we again applied KPFM to inspect respective surface potential distribution. Apparently, polarization reversal occurs only at $z^+$ surface under the application of the positive voltage pulses, while at $z^-$ surface of lithium niobate and $SiO_2$ only charge injection happens (Figure 3a,d,g). The profile of the surface potential has been fitted by the theoretical equation (Supporting Information S3) and average value of the charge density has been extracted and plotted versus amplitude of the applied voltage pulses (Figure 3).

First, the dependencies on duration and amplitude for all three materials were quite similar between each other (Figure 3) and, second, they resemble very closely the dependence of the domain area (Figures 1,2). Interestingly, the similar potential values measured for $SiO_2$ and lithium niobate indicate that injected charge contributes to reducing the electric field generated by the probe. The latter highlights again contact-limited injection dynamics. The dependence of the injected charge on amplitude of the voltage pulses measured in the $SiO_2$ corresponds to an exponent of $1.9 \pm 0.1$ (Figure 3c), apparent SCLC behavior. The dependence of injected charge on amplitude for $z^-$ surface of $LiNbO_3$ exhibits a power-law behavior, with an exponent of $\sim 1.4 \pm 0.1$ (Figure 3f). In the case of polarization switching in $LiNbO_3$ on the z- surface, the exponent for the dependence of injected charge on voltage is about $2.3 \pm 0.2$ (Figure 3i), which indicates the same SCLC current behavior. The exponent of injected charge dependence on amplitude was close to the domain area dependence in the same experiment: $2.2 \pm 0.1$ (Figure S4).

$SiO_2$ is well-known isolator with a band gap around 9 eV with a large density of the traps. However, in the surface layer they can be rapidly filled with injected charge of the probe with further establishing SCLC over filled traps and $m \approx 2$. At $z^-$ lithium niobate surface, the conditions differ from the $z^+$ surface. Recently, the (001) termination was simulated to exhibit predominantly p-type conductivity[42]. Alternatively, the negative bound charge can create specific conditions for the spreading of positively injected charge through band bending, in a manner similar to that observed at charged domain walls[43]. In the case of ferroelectric switching under application of a positive voltage to the $z^+$ surface, one can observe SCLC with deep traps mechanism, with an exponent of $m \approx 3$ at low voltages and $m < 3$ at higher voltages, where charge traps become partially filled by the injected carriers. Importantly, stable polarization switching is absent at 10 V, which leads to an apparent deviation in the charge behavior at this amplitude. It is also noteworthy that the amount of injected charge is similar at 10 V and 20 V, suggesting that part of the injected charge is compensated by stable domain formation. Thus, ferroelectric domain formation partially captures the injected charge, while further space-charge transport is governed by the filled traps. This may explain the difference between the exponents for the injected charge and the domain area, since in the case of polarization switching the total charge detected by KPFM is a combination of both the bound charge of the forming domain and the injected charge.

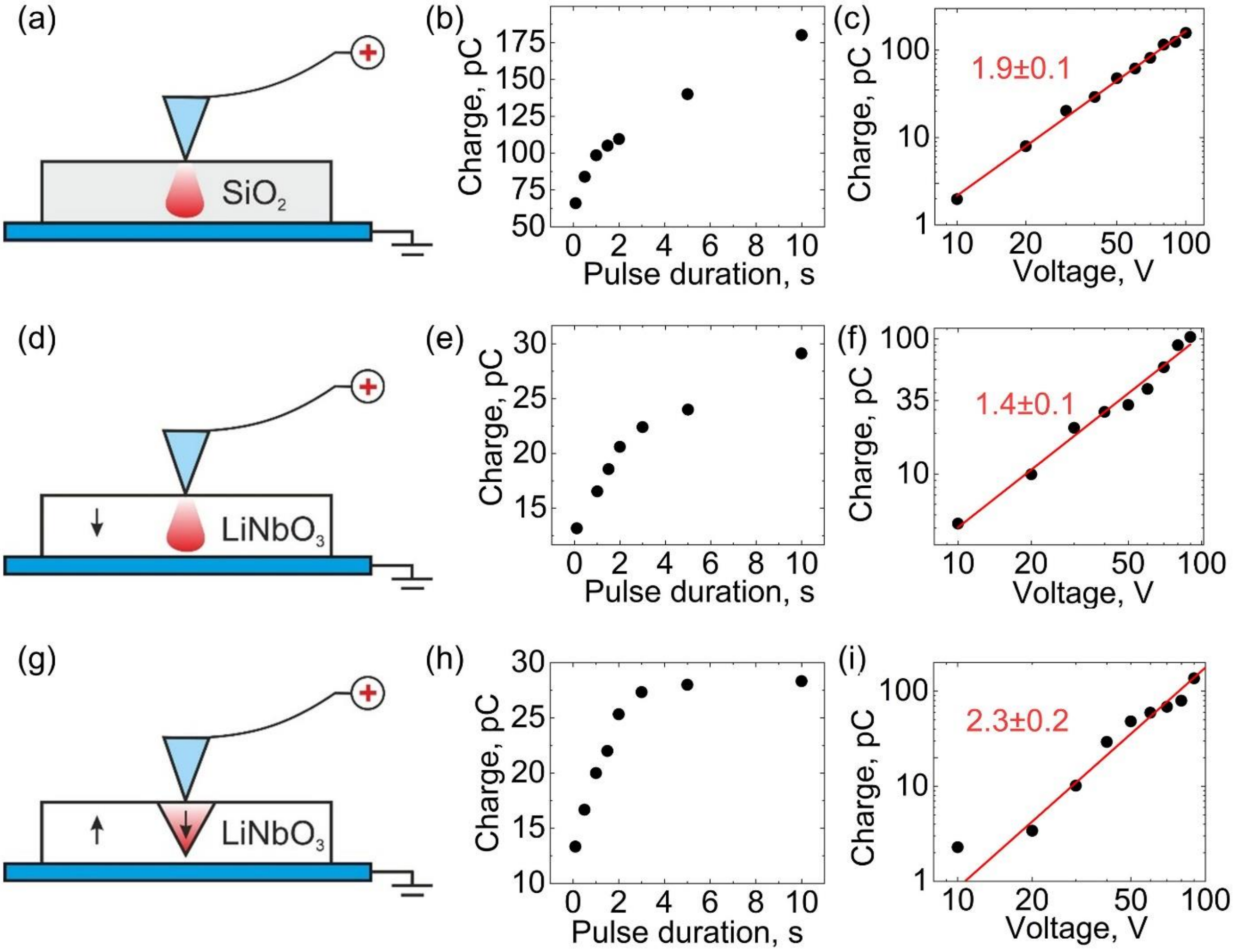

Figure 3. Dynamics of the charge injection into (a)-(c) $SiO_2$ and (d)-(i) $LiNbO_3$ with (d)-(f) downward and (g)-(i) upward direction of the polarization. (a), (d), (g) Schematics indicating conditions of the voltage pulse application. Arrow indicates direction of the spontaneous polarization. (b), (e), (h) Dependences of the injected charge on the pulse duration (at 100 V amplitude) and (c), (f), (i) pulse amplitude (at 10 s duration).

**3. Verification of the screening-charge limited mechanism**

SCLC behavior for the displacement current derived from the domain area assumes that charge injection from the SPM tip has a critical contribution to the switching compared to other mechanisms. To further clarify that, we performed additional experiments with the switching by the positive voltage pulses at the position of the domain wall (Figure 4). Comparison of the dependencies of the domain radius for the switching far from position of the domain wall (Figure 4c) and at the domain wall (Figure 4b), demonstrates a significant divergence of the formed domain radiuses (Figure 4f-h, Figure S5). However, the dependencies of the domain area perfectly overlapped, which have been repeatedly confirmed across a wide range of voltages and pulse durations (Figure 4i-k, Figure S5).

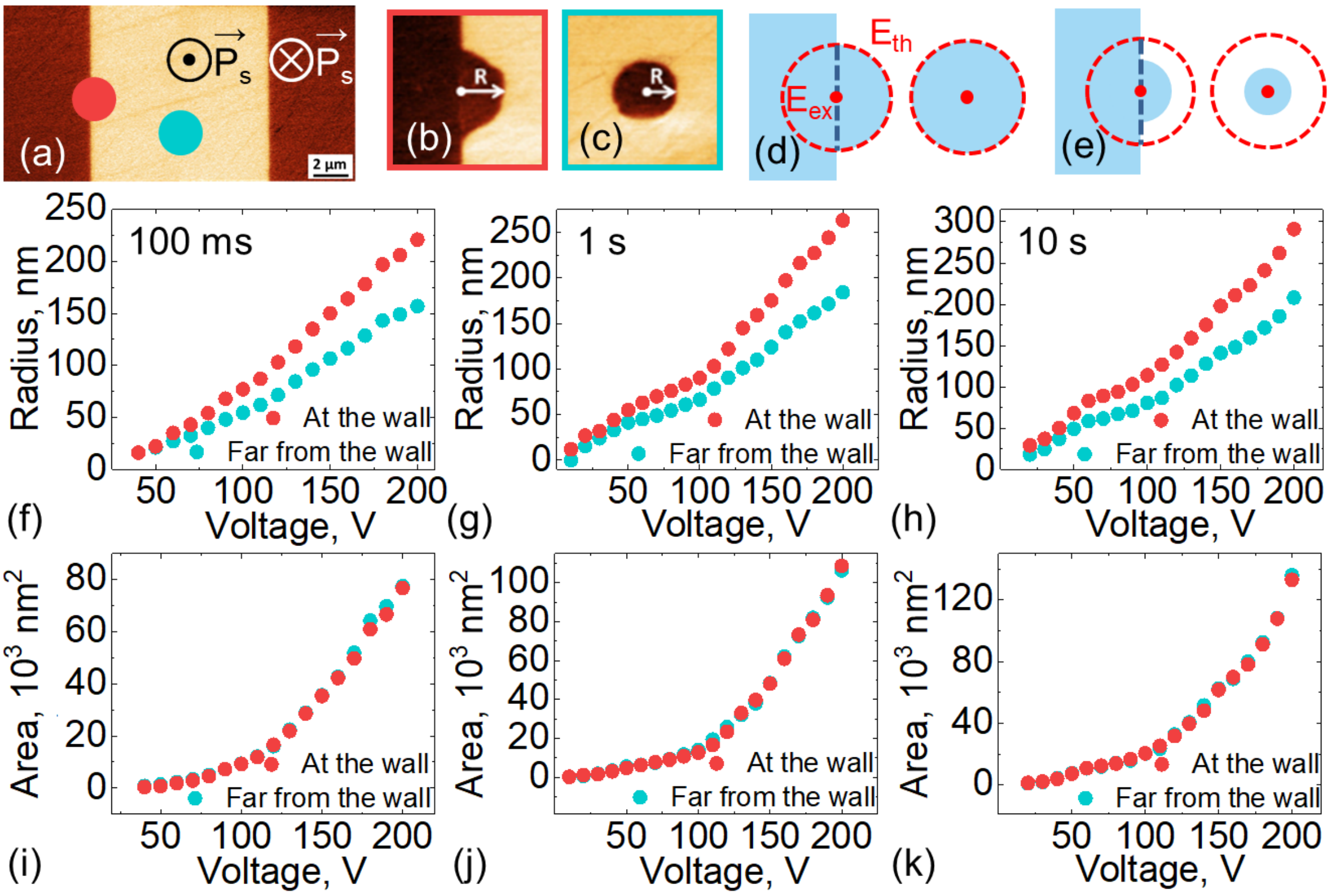

Figure 4. Local switching at the domain wall and far from the wall: (a) the domain structure of the sample: the red point is the point of application of an electric field for local switching on the domain wall, the cyan point is that for local switching on the domain wall; formed domains: (b) at the domain wall, (c) far from the domain wall; mechanisms of local switching: (d) voltage-limited regime, (e) current-limited regime; voltage dependencies of (f)-(h) domain radius and (i-k) domain area for pulse durations of: (f, i) 100 ms, (g, j) 1 s, (h, k) 10 s.

The overlapping of the domain areas in two cases indicates that charge injection is a dominant mechanism of the screening and limiting factor for the switching. If switching is limited by the electric field falling below the threshold required for polarization reversal at a certain radius, one would expect the domain radii to coincide, regardless of the location of the electric-field application. Typically, switching in a ferroelectric material occurs under the action of the polar component of the electric field. The total electric field driving switching at the domain wall can be written as follows:

$$E_{sw}^{z}(r,t) = E_{ex}^{z}(r) - E_{dep}^{z}(r,t) + E_{scr}^{z}(r,t) - E_{th}^{z}, \quad (2)$$

where $E_{ex}^{z}(r)$ – external applied electric field, $E_{dep}^{z}(r,t)$ and $E_{scr}^{z}(r,t)$ – temporally-dependent depolarization (produced by bound charges) and screening (produced by screening charges) fields, $E_{th}$ – threshold field of the polarization reversal. In the voltage-limited regime, we assume $E_{scr}^{z}(r,t)$ is always enough to screen $E_{dep}^{z}(r,t)$, making $E_{scr}^{z}(r,t) - E_{dep}^{z}(r,t) \approx 0$ at some moment of time, while the limiting domain radius, $r_{lim}$, is determined by the distance $r$, where $E_{ex}^{z}(r) > E_{th}^{z}$. In the screening-current-limited regime, the applied electric field is assumed to be always sufficient to switch the polarization, i.e., $E_{ex}^{z}(r) > E_{th}^{z}$. However, polarization switching only occurs in the region where $E_{scr}^{z}(r,t) - E_{dep}^{z}(r,t) \approx 0$, i.e. the maximal radius is determined solely by the distance at which enough screening charge from the tip has reached. The latter also implies a quasi-static regime for both the domain-radius dependence on pulse duration and that on the amplitude of the applied switching pulses. The duration dependence cannot be directly interpreted as reflecting the time evolution of domain-wall motion under an external electric field, rather, domain-wall motion is governed by the established regime of the charge transport.

To further confirm our conclusions, we used external resistance in the circuit aiming to limit injection (screening) current (Figure 5a). One can expect that applied voltage would be redistributed between the external resistance and tip-surface gap according to relation:

$$U_{ex} = IR_{ex} + U_{gap}, \quad (3)$$

where $I$ – total screening current and $U_{ex}$ is an applied external voltage. In the gap, it can be written through the SCLC equation (1) exchanging $U$ to $U_{gap}$. To simplify and get analytical solution, let assume $m = 2$. Other exponent values lead to transcendent equations with solely numerical solution but with a not strong difference in the final dependence. Substituting (1) to (2) gives:

$$U_{ex} = \sqrt{\frac{I}{a}} + IR_{ex}. \quad (4)$$

Using further substitution $x = \sqrt{\frac{I}{a}}$, one can derive: $x + R\sqrt{a}x^2 = \sqrt{a}U$. Solving this square equation, final relation between screening current and value of the external resistance, $R_{ex}$, can be achieved:

$$I = \frac{1}{a} \cdot \left(\frac{-1+\sqrt{1+4aR_{ex}U_{ex}}}{2R_{ex}}\right), \quad (5)$$

As the domain area is proportional to the total injected charge, one can assume domain area to be proportional to $I$, as well. Thus, when including external resistance, a limitation of the screening current and a decrease in the area of the forming domain is expected. One should note that this limitation is possible only in the case of the current-limited switching. Field-limited switching is apparently should not "feel" the external resistance. We also should note here that the total capacitance of the system mainly comes from the sample and can be estimated in our case to be 0.2-0.5 nF, then assuming lowest $\tau = R_{ex}C$ to be 0.5 s for 1 GOhm, which cannot significantly change the shape of the voltage pulses for the used pulse duration. At 10 GΩ, approximately 13.5% of the charge is lost due to the $R_{ex}C$ charging effect, however, this does not substantially affect the overall shape of the curve in Figure 5b.

One could see that equation (4) well describes the experimental data (Figure 5b), thereby confirming screening current-limited mechanism of the local switching. $R_{ex}$variation allows to control the average domain size, which can be used to precisely control the domain size in the micro- and nanodomain engineering applications.

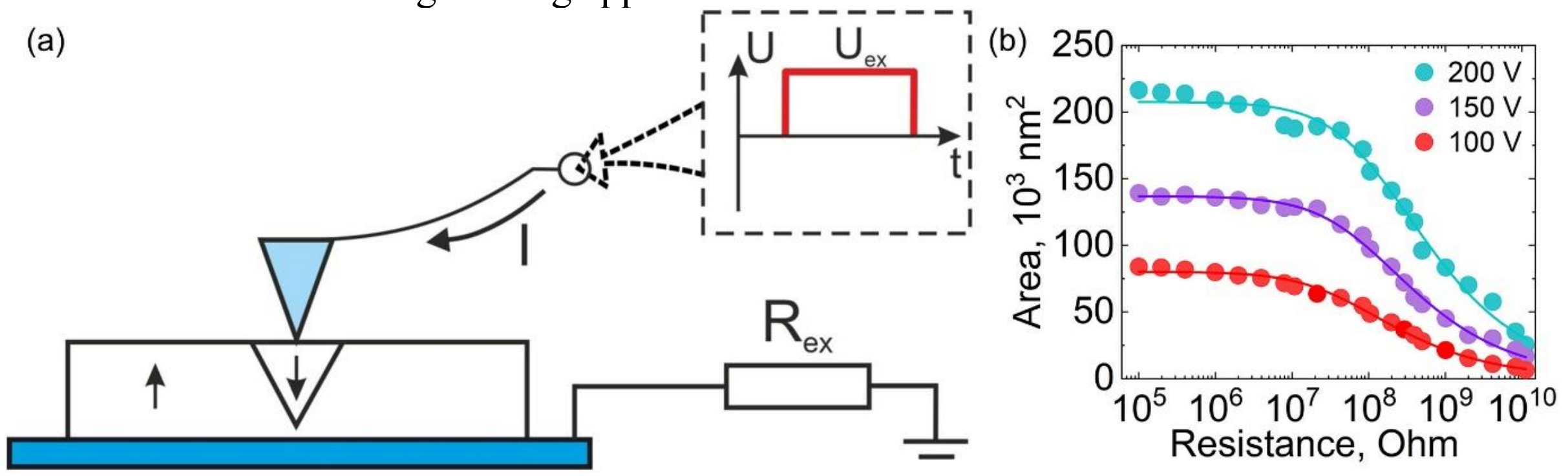

Figure 5. (a) Schematic illustration of used methodology addition of an external resistance; (b) Dependence of domain area from RC circuit resistance.

**4. The simulations of the screening-limited switching dynamics**

Finally, finite-element simulations were used to corroborate the experimental trends and to gain insight into the underlying processes. Charge injection was implemented by applying an ohmic current-flow boundary condition over the probe-sample contact, with the contact resistance, $R_c$, set empirically in the range 10 kΩ to 1 MΩ. The charge injection, spreading and decay has been simulated using drift-diffusion equations:

$$\frac{\partial\sigma}{\partial t} = \nabla \cdot (D\nabla\sigma - \mu_e E\sigma), \quad (6)$$

The first term of the equation, $D\nabla\sigma$, describes diffusion-driven charge spreading from regions of high charge density ($\sigma$) to regions of lower charge density, where $D$ is the diffusion coefficient. The second term, $-\mu_e E\sigma$, represents the drift of charge carriers under the electric field, $E$, with $\mu_e$ being electron mobility in the material, while the divergence operator accounts for the local redistribution of charge density in space.

To describe the flux of charge carriers across the surface the following boundary condition was implemented:

$$\boldsymbol{n} \cdot (D\nabla\sigma - \mu_{ch} E\sigma) =\cdot (\boldsymbol{n} \cdot \boldsymbol{E})/R_c, \quad (7)$$

where $\boldsymbol{n}$ is the unit vector normal, $R_c$ is a contract resistance. This equation describes charge-carrier injection from the conductive tip into the sample. It expresses a balance between carrier transport within the material, governed by diffusion and drift, and carrier injection across the tip–sample interface, controlled by the local electric field and the interfacial resistance.

To simplify the model, we assumed (1) an ohmic injecting contact and (2) no charge trapping, i.e. we did not include microscopic electronic-structure effects. As a result, the model is not intended to reproduce the absolute experimental currents, but it captures the qualitative features of charge accumulation and its impact on domain growth, thereby supporting the proposed mechanism. Polarization switching was modeled by considering domain growth as a variation of the bound charge on the polar surface. The bound surface charge was simulated as a 5 nm thick surface layer, with the charge distributed in depth according to a truncated Gaussian profile. The domain was represented by a sigmoid profile describing the polarization transition from $-P_s$ to $P_s$:

$$P(x,t) = -P_s + \frac{2P_s}{1+exp\left[-\frac{x-x_{DW}(t)}{w}\right]}, \quad (8)$$

where $x_{DW}(t)$ – is the domain wall position and $w$ – is an effective domain wall width. The simulations were initialized in a single-domain state. Domain-wall motion was simulated using a simple linear mobility law:

$$v = \mu_{DW}(E - E_{th}), \quad (9)$$

where $\mu_{DW}$ is domain wall mobility. When the local field exceeded $E_{\text{th}}$, polarization reversal was allowed, producing a local change $\Delta P = 2P_s$ and the associated bound charge. Domain wall dynamics and injected charge motion redistribution were solved as a fully coupled time domain problem.

For comparison, we also analyzed a simplified case without domain formation to isolate charge-injection dynamics (Supporting Information 5). In this case, injected charge spreads isotopically through the bulk (Figure S6a, Video S1), and the current–voltage response follows a Child–Langmuir law consistent with a space-charge-limited current (SCLC) regime ($m \approx 3/2$, Figure S6b). The total injected charge increases approximately logarithmically with time, in agreement with experiment (Figure S6b). The charge drops significantly once $R_c$ approaches roughly $10^7 \div 10^8$ Ohm (Figure S7a), indicating a change in the charge-spreading behavior between bulk and surface pathways. An increase in mobility results in a larger total amount of injected charge while the overall charge-transport behavior remains similar (Figure S7b).

Introducing a domain growth, as described above, modifies the injected-charge distribution by redirecting a fraction of carriers toward screening the bound charge associated with the newly formed domain (Video S2). The resulting switching dynamics proceed through a series of characteristic stages (Figure 6, Figure S8): (1) charge injection beneath the probe, with isotropic charge spreading similar to the case without domain formation (uniform area in Figure 6a), (2) domain nucleation, accompanied by a corresponding localization of charge density within the growing domain region (Figure 6a); (3) lateral expansion of the domain and the associated space-charge region until its extent approaches the probe-surface contact radius (Figure 6b, c); (4) once the domain radius exceeds the contact size, further growth becomes limited by the screening current, leading to a near-logarithmic evolution of the domain radius with time, consistent with the experimental kinetics (Figure 6d-f); and (5) at long pulse durations, the system approaches the threshold voltage for polarization reversal and the domain no longer propagates.

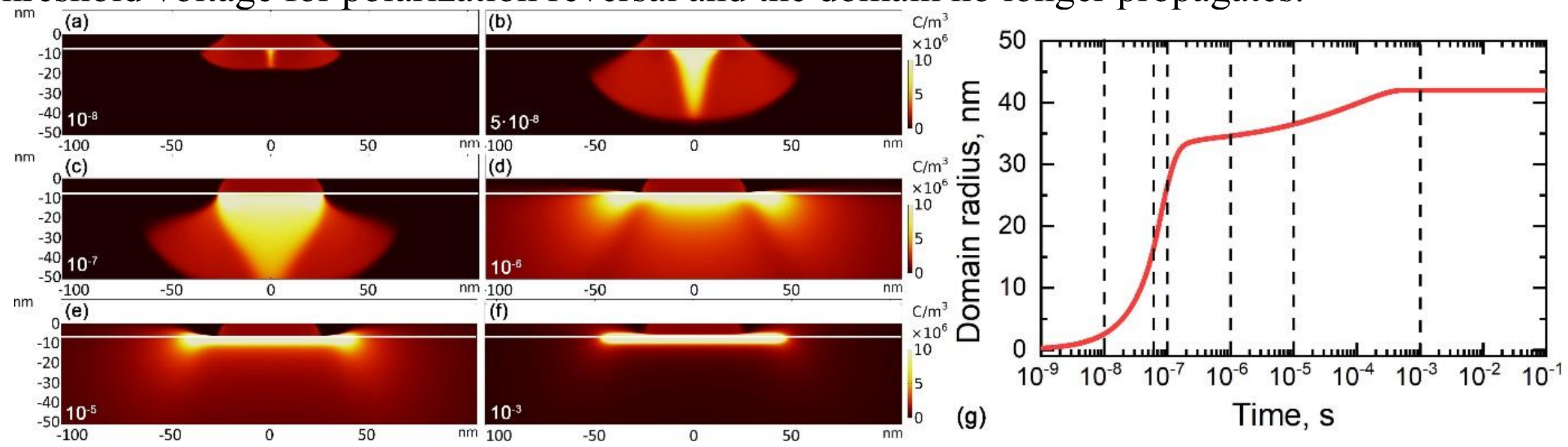

Figure 6. (a)-(f) Spatial distribution of the charged density beneath the biased probe at four representative moments of time during the polarization switching process and (g) respective temporal evolution of the domain radius. $\mu_e = 10^{-9}$ m$^2$/(V·s), $\mu_{DW} = 10^{-10}$ m$^2$/(V·s).

Varying the simulation parameters helps clarify the kinetic regimes governing domain evolution. Increasing the domain-wall mobility leads to larger domain sizes (Figure S9), while increasing the bulk charge-carrier mobility has a non-monotonic effect on the domain size (Figure 6, Figure S10a). The latter parameter largely determines the growth kinetics in the current-limited regime: the faster the injected charge propagates into the bulk, the weaker the lateral charge spreading, which in turn suppresses lateral domain-wall motion (Figure 6, "leakage-current limited regime"). However, at low carrier mobilities, where charge leakage into the bulk is minimized, the behavior is reversed (Figure 7, "injection-current-limited regime", Figure S11, Video S3). In this case, a larger fraction of the injected charge contributes directly to domain-wall motion, resulting in enhanced domain growth while $\mu_e$ regulate screening charge supply rate. In practice, the effect of the leakage of the carriers in the bulk can be even stronger if conductivity along the domain wall is taken into account[34], as it can introduce vertically enhanced conduction and thereby influence the final domain shape[28]. Thus, the mobility of charge carriers in the material bulk governs the transition between injection-current-limited and leakage-current-limited growth regimes. Finally, simply increasing the threshold voltage mainly causes an earlier saturation of domain growth (Figure S10b). Overall, relative changes in charge-carrier mobility and domain-wall mobility modify domain size while preserving an approximately logarithmic time dependence (Figure S9).

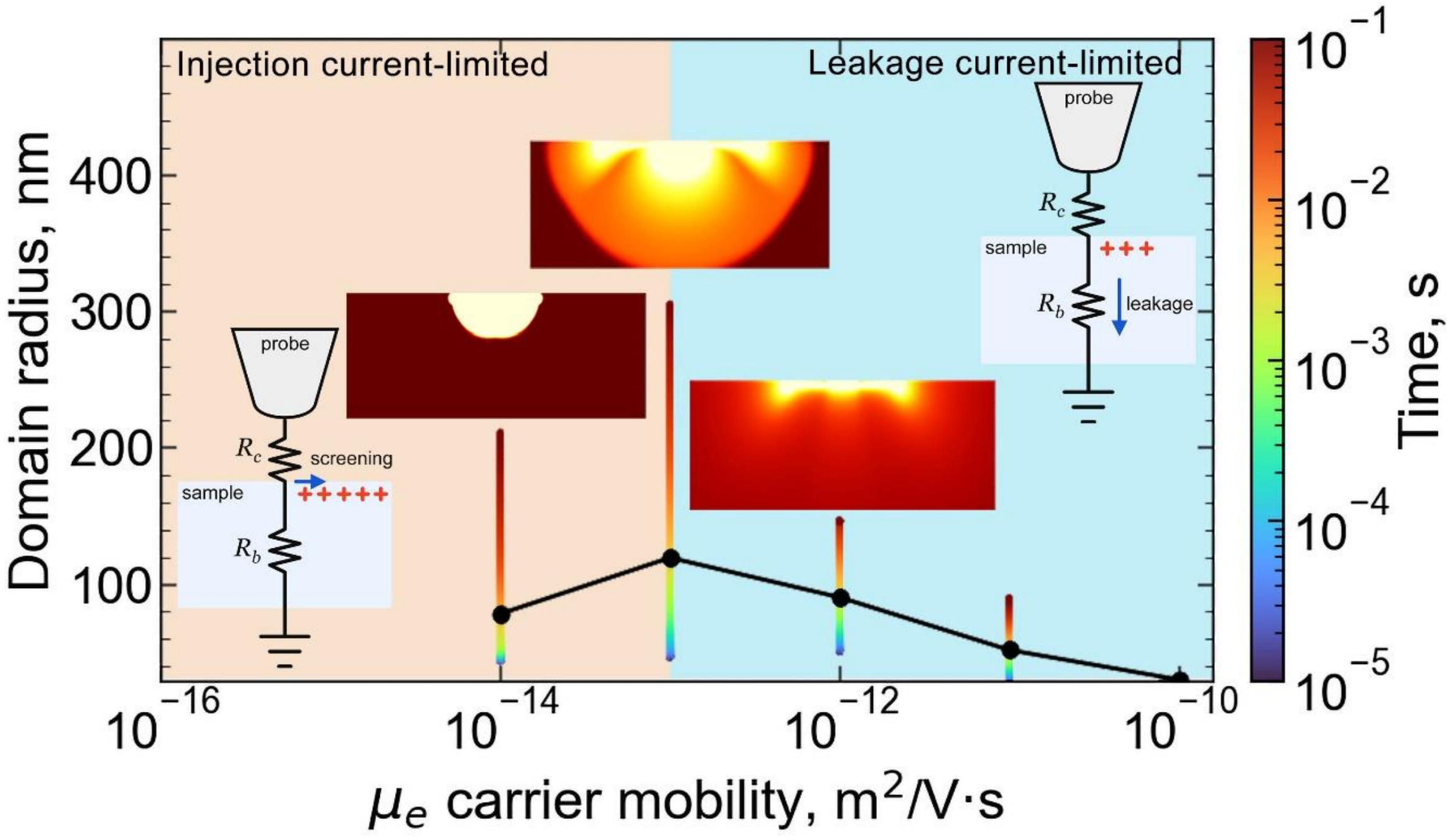

Figure 7. Domain radius as a function of charge-carrier mobility $\mu_e$ in the bulk. Individual points correspond to domain radii calculated at different times in the range from $10^{-5}$to $10^{-1}$s, with color indicating time on a logarithmic scale. Black circles connected by a solid line denote the average domain radius over the considered time interval. Images correspond to t = 100 ms. The schematics illustrate the two charge-injection regimes: dominant leakage channel through the bulk resistance, $R_b$, or dominant screening of the polarization bound charge by the injected charge, $\mu_{DW} = 10^{-11}$ m$^2$/(V·s).

Importantly, in the beginning stage of the domain growth, the dependence of the domain radius on time is nearly linear in time (Figure S12), consistent with a field-limited regime governed by the kinetic law given in Eq. 9 (Supplementary Information 1, Equations S11, S12). In practical terms, this means that, immediately beneath the probe, a voltage-limited regime can be achieved for short voltage pulses. At present, this regime is difficult to access in conventional SSPFM measurements because of limited current and displacement signal sensitivity. However, we would expect it to be probed more effectively using interferometric PFM[44] or scanning microwave microscopy[45] at low applied DC or AC voltages.

## 5. CONCLUSIONS

Polarization switching is commonly described using thermodynamic nucleation models, thermally activated domain-wall motion, or creep-like kinetics. However, such models often fail to fully capture the observed switching kinetics and may yield inconsistent physical parameters. Here, we show that local polarization switching under an SPM tip is controlled primarily by charge transport, rather than by the tip-induced electric-field distribution and domain-wall kinetics alone. Specifically, local ferroelectric switching in lithium niobate is shown to enter a current-limited regime controlled by charge injection, screening-charge relaxation, and space-charge-limited-current transport. The observed scaling of domain size with voltage amplitude and pulse duration is consistently explained by screening-charge dynamics, indicating that the apparent switching kinetics are dictated by transport and relaxation processes. Importantly, this conclusion is not restricted to tip-induced switching. Whenever polarization reversal involves significant charge injection, field redistribution across the interface, or ionic dynamics, the apparent switching kinetics and hysteresis behavior may be dictated by charge transport. Therefore, reliable extraction of ferroelectric switching parameters, whether in local probe experiments or capacitor geometries, requires explicit identification of the active switching regime: field-limited or transport/current-limited.

Beyond interpretation, this regime-based perspective provides a practical control parameter for nanoscale ferroelectric engineering. Since domain growth becomes limited by the screening current, regulation of the current enables deterministic control over nanoscale ferroelectric domain size and shape. More broadly, adopting a regime-aware framework that explicitly accounts for screening and transport processes expands the understanding of polarization switching and opens new opportunities for controlled domain studies, domain patterning, and fabrication of functional ferroelectric nanostructures.

## ACKNOWLEDGMENTS

The research was made possible by the financial support of the Ministry of Science and Higher Education of the Russian Federation (state task FEUZ-2023-0017).

Part of this work (DA) was developed within the scope of the project CICECO-Aveiro Institute of Materials, UID/50011/2025 (DOI 10.54499/UID/50011/2025) & LA/P/0006/2020 (DOI 10.54499/LA/P/0006/2020), financed by national funds through the FCT/MCTES (PIDDAC). DA acknowledges support by the individual contract 2023.09305.CEECIND/CP2840/CT0020 through contract (DOI: 10.54499/2023.09305.CEECIND/CP2840/CT0020) through national funds provided by FCT − Fundação para a Ciência e a Tecnologia.

## SUPPORTING INFORMATION

*S1. The calculations of the domain-radius dependencies on the time and amplitude of the applied voltage pulses in electric field-limited models.*

The lateral growth of a ferroelectric domain created by a biased scanning probe tip can be described by considering the motion of the domain wall driven by the local electric field produced by the tip. In the field-limited regime of polarization reversal, we assume that domain nucleation under the tip is controlled mainly by the spatial distribution of the polar electric-field component. For a biased probe, this component can be written as[1]:

$$E_z(r) = \frac{1}{4\pi\varepsilon_0}\frac{2CV_{tip}}{\varepsilon+1}\frac{R_{tip}}{\left(r^2+\mathrm{R}_{tip}^2\right)^{3/2}}, \tag{S1}$$

where $r$ is the lateral distance from the contact point, $C$ is the effective tip–sample capacitance, $\varepsilon$ is the dielectric permittivity of the sample, $\varepsilon_0$ is the vacuum permittivity, $V_{\text{tip}}$ is the voltage applied to the tip, and $R_{\text{tip}}$ is the tip radius. Thus, the polar field component decays approximately as $E_z \propto r^{-3}$at large $r$.

The velocity of the domain wall is assumed to follow a linear kinetic law with a threshold field:

$$v = \mu_{DW}(E - E_{th}), \tag{S2}$$

where $\mu_{\text{DW}}$ is the domain-wall mobility and $E_{\text{th}}$ is the threshold field for the polarization reversal, and domain walls don't move under the field below $E_{\text{th}}$. In

$$dR/dt = v(r) = \mu_{DW}(E_z(r) - E_{th}). \tag{S3}$$

Introducing a constant:

$$A = \frac{1}{4\pi\varepsilon_0}\frac{2CV_{tip}R_{tip}}{\varepsilon+1}, \tag{S4}$$

and substituting the electric field expression into the kinetic equation yields the differential equation governing the domain radius evolution:

$$dR/dt = \mu_{DW}\left(\frac{A}{\left(r^2+\mathrm{R}_{tip}^2\right)^{3/2}} - E_{th}\right). \tag{S5}$$

Since the electric field decreases monotonically with increasing distance from the tip, the domain wall velocity also decreases during lateral domain growth. The time dependence of the domain radius is therefore determined by the integral form of the kinetic equation:

$$t(R) = \frac{1}{\mu_{DW}}\int_0^R \frac{dr}{\left(\frac{A}{\left(r^2+\mathrm{R}_{tip}^2\right)^{3/2}} - E_{th}\right)}. \tag{S6}$$

This representation makes it possible to clarify the behavior of the solution in two important limits. Far from the limiting radius, when the local field substantially exceeds the threshold field, i.e. $E_z(r) \gg E_{th}$, the threshold term can be neglected in the denominator. In this regime,

$$t(R) = \frac{1}{\mu_{DW}A}\int_0^R \left(r^2 + \mathrm{R}_{tip}^2\right)^{3/2} dr. \tag{S7}$$

This integral can be evaluated analytically,

$$\int_0^r \left(r^2 + R_{tip}^2\right)^{3/2} dr = \frac{r(2r^2+5R_{tip}^2)\sqrt{r^2+R_{tip}^2}}{8} + \frac{3R_{tip}^4}{8} \ln\left(\frac{r+\sqrt{r^2+R_{tip}^2}}{R_{tip}}\right) + C, \quad \text{(S8)}$$

which yields:

$$t = t_0 + \frac{1}{8\mu_{DW}A} \left| \frac{r(2r^2+5R_{tip}^2)\sqrt{r^2+R_{tip}^2}}{8} + \frac{3R_{tip}^4}{8} \ln\left(\frac{r+\sqrt{r^2+R_{tip}^2}}{R_{tip}}\right) \right|_0^R . \quad \text{(S9)}$$

While further inverse transformation of the equation (S9) to $R(t)$ dependency in an analytical form is not possible, it seems useful to consider different limits in the field dependence given by equation S1.

At small radii, $r \ll R_{\text{tip}}$, the electric field is approximately constant,

$$E_z \approx \frac{A}{R_{tip}^3}, \quad \text{(S10)}$$

so that the integral S3 can be easily integrated given domain wall velocity as well approximately constant and the radius grows linearly with time and tip voltage,

$$R(t) \approx \mu_{DW} \frac{A}{R_{tip}^3} t, \quad \text{(S11)}$$

$$R(t) \propto V_{tip}. \quad \text{(S12)}$$

At larger radii, but still in the regime $E_z(r) \gg E_{th}$, the field decays approximately as:

$$E_z \approx \frac{A}{r^3} \quad \text{(S13)}$$

The kinetic equation then becomes:

$$dR/dt = \mu_{DW} \frac{A}{r^3}, \quad \text{(S14)}$$

which after integration gives:

$$r(t) \approx (4\mu_{DW}At)^{\frac{1}{4}}, \quad \text{(S15)}$$

$$R(t) \propto \left(V_{tip}\right)^{\frac{1}{4}}. \quad \text{(S16)}$$

A qualitatively different behavior emerges near the limiting radius. Let $r = r_{lim}$ be the point at which the local field becomes equal to the threshold field. Expanding the denominator of the time integral near $r_{lim}$, one obtains

$$E_z(r) - E_{th} \approx \left.\frac{dE_z}{dr}\right|_{r_{lim}} (r - r_{lim}). \quad \text{(S17)}$$

Since $dE_z/dr < 0$, the denominator decreases linearly to zero as $r \to r_{lim}$. Therefore, in the vicinity of $r_{lim}$,

$$t(R) \sim \frac{1}{\mu_{DW}} \int_0^R \frac{dr}{r - r_{lim}}. \quad \text{(S18)}$$

This integral diverges logarithmically. Physically, this divergence means that the domain wall velocity continuously decreases as the local field approaches the threshold value and becomes exactly zero at $r = r_{lim}$. As a result, the limiting radius is approached only asymptotically and cannot be reached within a finite time in the deterministic linear-threshold model. The absence of convergence of the time integral is a direct consequence of the linear vanishing of the driving force $E_z(r) - E_{th}$ at the stopping point.

The limiting domain radius is obtained from the condition of zero domain wall velocity,

$$dR/dt = 0, \quad \text{(S19)}$$

which is equivalent to

$$E_z(r_{lim}) = E_{th}. \quad \text{(S20)}$$

Substituting the field distribution yields:

$$\frac{A}{\left(r_{lim}{}^2 + R_{tip}^2\right)^{\frac{3}{2}}} = E_{th}. \quad \text{(S21)}$$

Solving for $r_{lim}$ gives:

$$\left(r_{lim}{}^2 + R_{tip}^2\right)^{\frac{3}{2}} = \frac{A}{E_{th}}, \quad \text{(S22)}$$

$$r_{lim}{}^2 + R_{tip}^2 = \left(\frac{A}{E_{th}}\right)^{\frac{2}{3}}. \quad \text{(S23)}$$

Therefore, the limiting radius is:

$$r_{lim} = \sqrt{\left(\frac{A}{E_{th}}\right)^{\frac{2}{3}} - R_{tip}^2}. \quad \text{(S24)}$$

Restoring the definition of $A$, the final expression can be written as

$$r_{lim} = \sqrt{\left[\frac{1}{4\pi\varepsilon_0} \frac{2CV_{tip}}{\varepsilon+1} \frac{R_{tip}}{E_{th}}\right]^{\frac{2}{3}} - R_{tip}^2}. \quad \text{(S25)}$$

This result shows that the maximum domain size is determined by the balance between the spatial decay of the tip-induced electric field and the threshold field required for domain wall motion. The domain radius initially increases with a finite velocity, then progressively slows down, and finally approaches the limiting value asymptotically, while the exact attainment of this radius is prevented by the logarithmic divergence of the corresponding time integral.

In capacitor geometry, domain-wall velocity often does not follow a linear relationship with electric field and it is more accurately described by an activated (Merz) law [2,3]:

$$v(E) = v_0 \exp\left(-\frac{E_{act}}{E}\right). \quad \text{(S26)}$$

This model does not include a strict threshold field, implying that domain growth could occur even at arbitrarily low fields. For lithium niobate family of the crystals, this clearly contradicts experimental observations[45,6], where domain growth is effectively suppressed below a relatively high threshold field. A common explanation is that additional internal fields contribute to the effective threshold and can pin the domain wall until the applied electric field exceeds this combined barrier $E_{th}$[7,8].

Therefore, equations S19-20 still applies and must be met in this case. The relationship between the limiting radius and the applied voltage exhibits minimal sensitivity to variations in the selected kinetic law. This follows from the fact that the saturation condition is set by the electrostatic requirement, rather than by the explicit form of $v(E)$. Because the tip-induced field scales linearly with $V_{tip}$, the position at which the local field drops to the threshold value is determined by essentially the same condition for both the linear and activated kinetic models. Consequently, the asymptotic voltage scaling of the limiting radius remains unchanged, $r_{lim} \propto V_{tip}^{1/3}$, whereas the main effect of the activated law is to modify the temporal approach to this limit. In particular, the near-threshold dynamics becomes much slower than in the linear case, although the final voltage-controlled limiting radius itself is nearly unaffected.

*S2. Dependences of the domain area and radius on the pulse duration for the different amplitudes of the applied voltage*

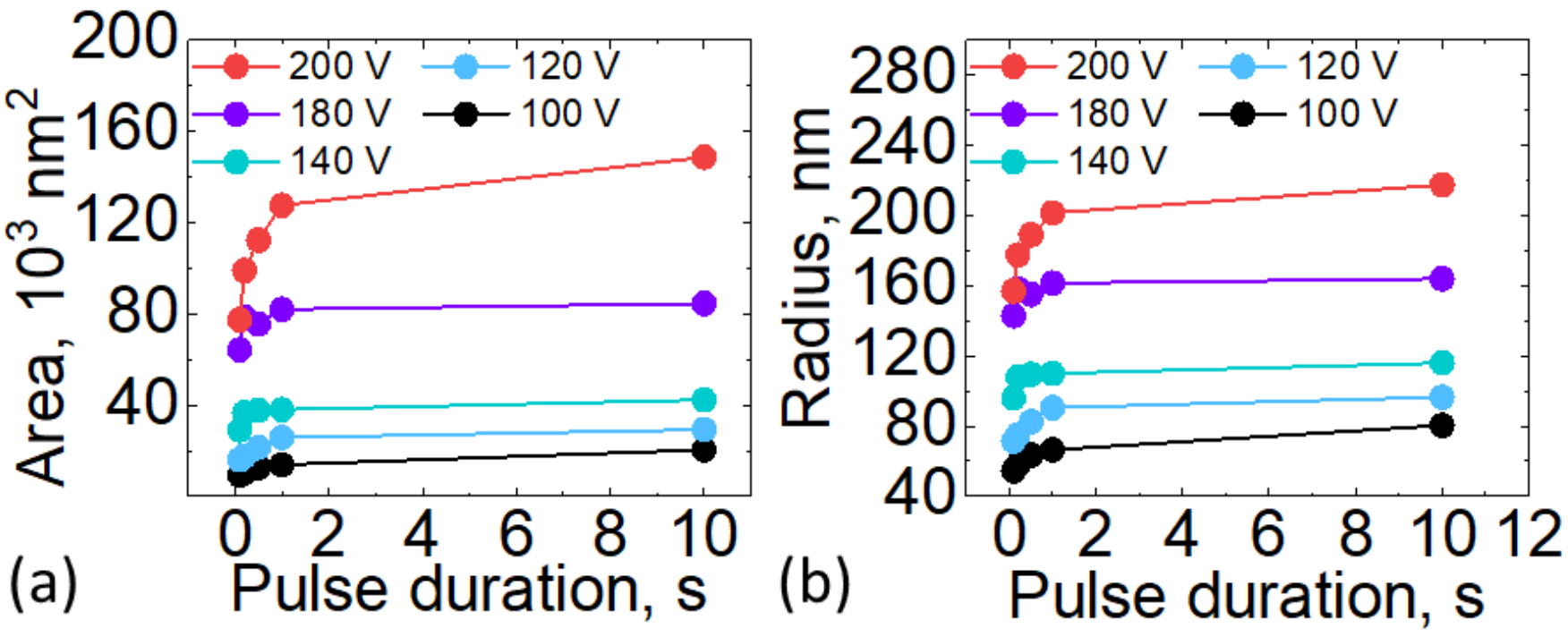

Figure S1. Dependencies of the domain (a) area and (b) radius on the pulse duration for the different amplitudes of the applied voltage.

*S3. Local hysteresis loop measurements*

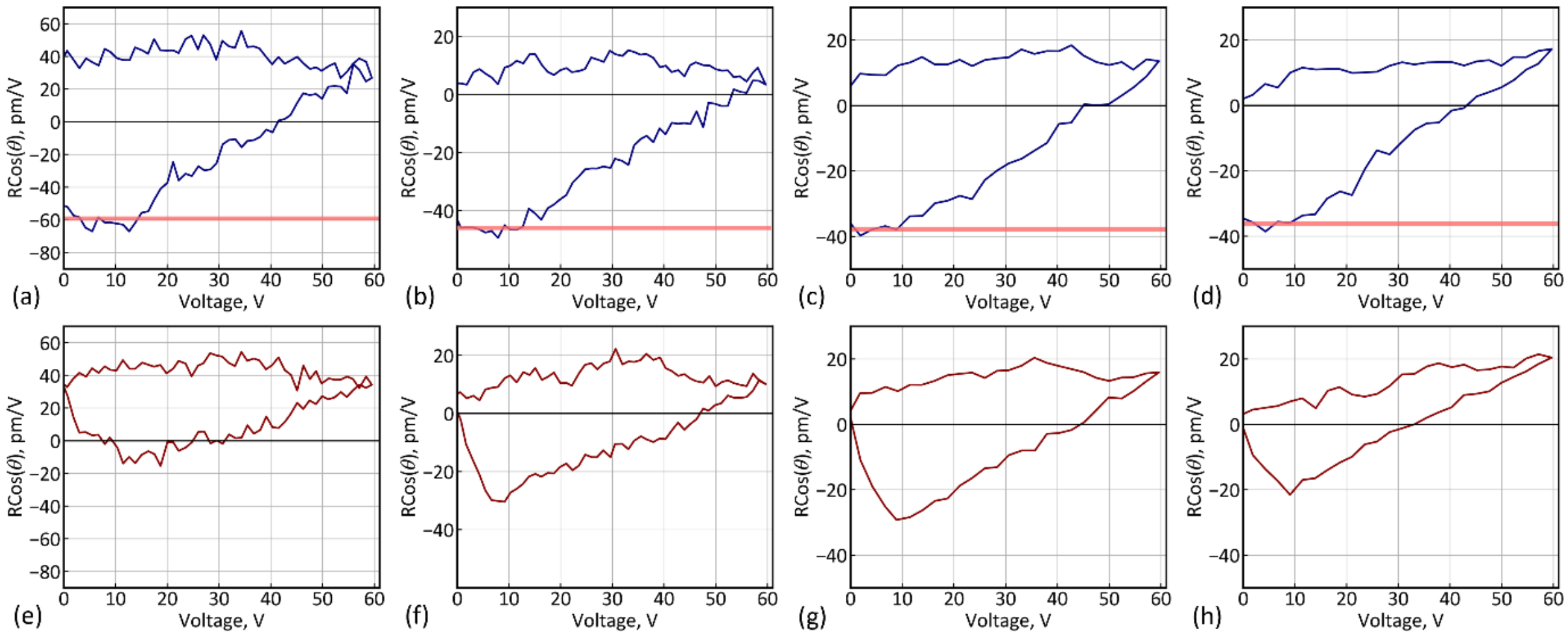

Figure S2. Piezoresponse dependencies on the applied DC voltage extracted from PUND SSPFM measurements. (a)-(d) P-N and (e)-(f) U-D branches of PUND-SSPFM curves for the different duration of the pulses in the PUND waveform: (a),(e) 10 ms; (b), (f) 100 ms; (c), (g), 500 ms; (d), (h) 1 s. The red line marks where the piezoresponse change exceeds the standard deviation, considered as the threshold voltage.

*S3. Approximation of surface potential profiles*

To interpret the KPFM image data, line profiles across the injected-charge region were extracted and analyzed (Figure S3). The total injected charge was estimated using an analytical model assuming an axisymmetric Gaussian surface-charge distribution. Within the 2D approximation, the surface potential profile was described by the corresponding electrostatic solution[9], which yields the characteristic Bessel-Gaussian form:

$$V(r) = \frac{Q}{(2\pi\varepsilon\varepsilon_0 a)} \cdot I_0\left(\frac{(r-r_0)^2}{2a^2}\right) \cdot \exp\left(-\left(\frac{(r-r_0)^2}{2a^2}\right)\right) + V_{SP0}, \quad \text{(S27)}$$

where *V(r)* is amplitude of surface potential, $Q$ – total injected charge, $\varepsilon$ – dielectric constant, $\varepsilon_0$ – dielectric constant of vacuum, $a$ – the width of the charge distribution, $I_0$ – is the zero-order modified Bessel function of the first kind, $r_0$ – injection center position, $r$ – radius of the injected charge area, $V_{SP0}$ – a background surface potential.

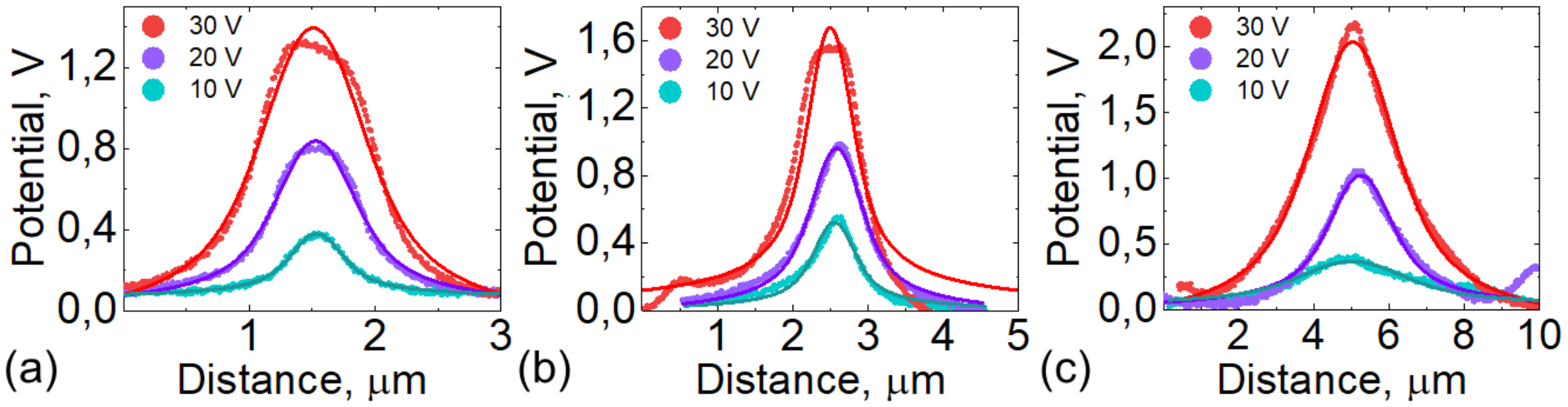

Figure S3. Approximation results of potential distribution profiles using equation 8 for: (a) $SiO_2$; $LiNbO_3$ with (b) downward and (c) upward direction of the polarization.

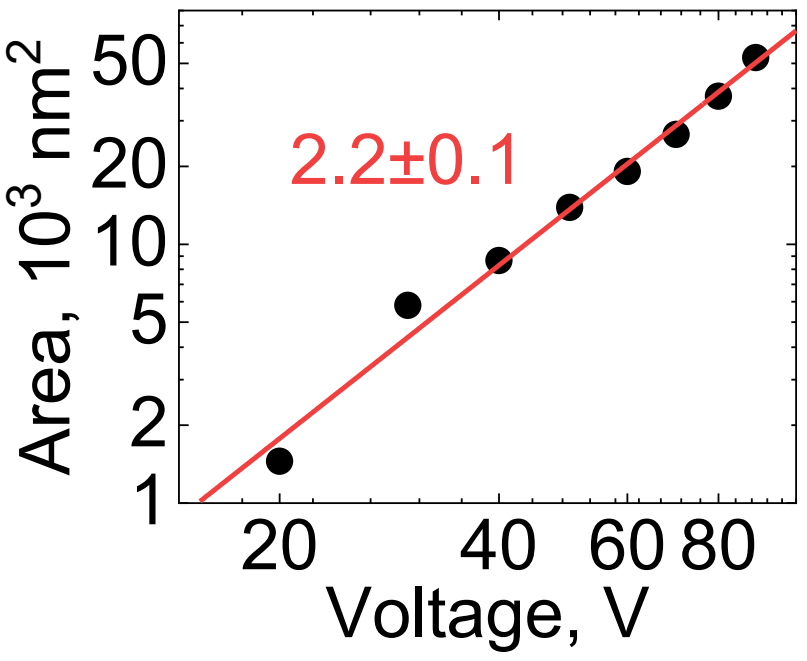

Figure S4. Dependence of the domain area on the pulse amplitude, corresponding to the charge dependence shown in Figure 3i of the main text.

*S4. Determination of the exponents of degrees characteristic of voltage dependence of the domain area*

To confirm the assumption of the charge-limited mechanism, a local switching experiment was conducted over a wide range of pulse amplitudes and durations. The obtained dependencies of domain radius on applied voltage show significant divergence, while the area dependencies coincide. These results are reproducible over a wide range of pulse durations (Figure 4, Figure S5).

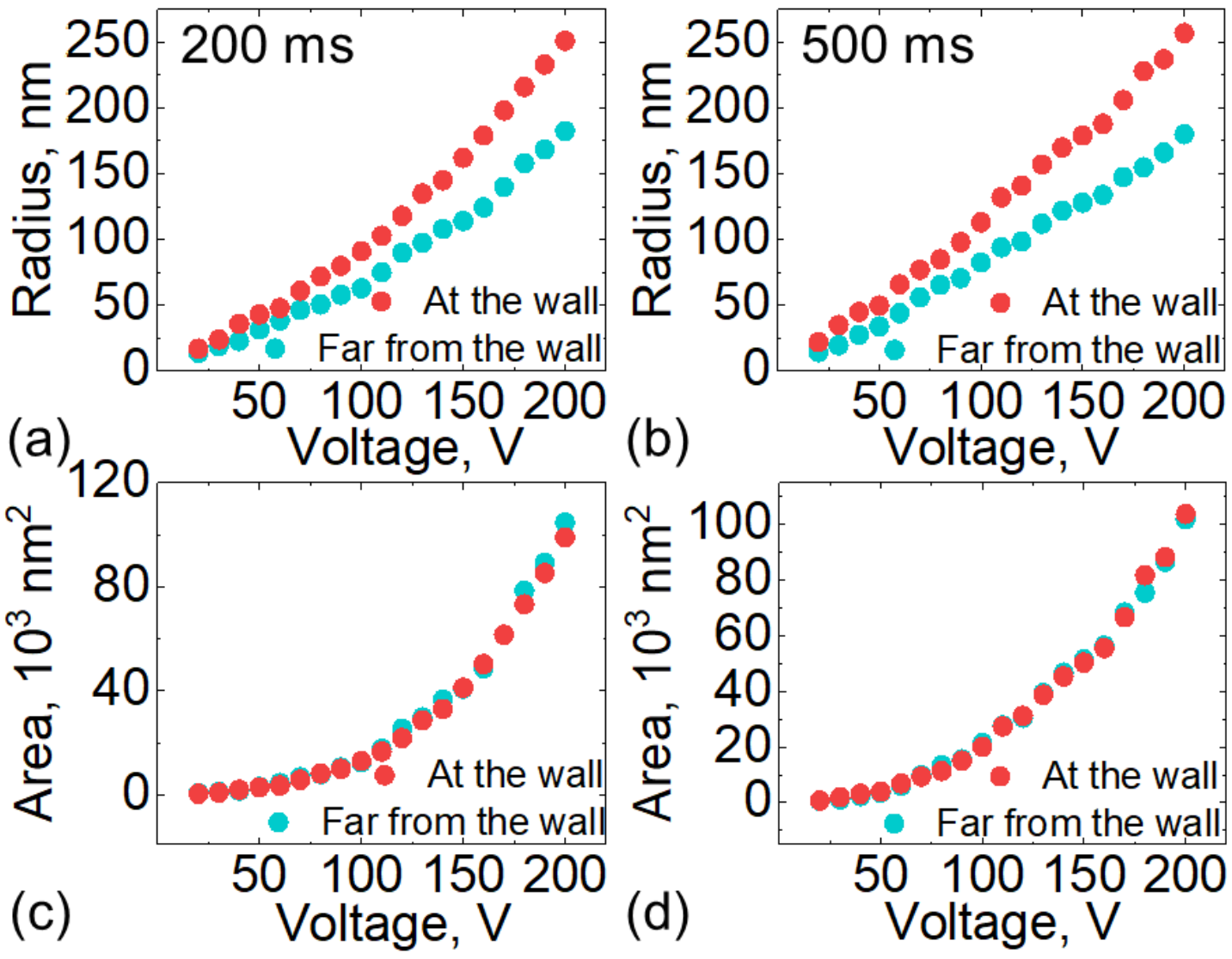

Figure S5. Dependencies of domain radius and domain area on the applied voltage for pulse durations of: (a), (c) 200 *ms*, (b), (d) 500 *ms*.

Exponent parameters were extracted for various voltage pulse durations and amplitude ranges (see Table S1).

Table S1. Exponents characterizing the relationship between domain area and applied voltage.

<table>
<tr><th></th><th colspan="2">Far from wall</th><th colspan="2">On wall</th></tr>
<tr><td>Pulse duration, s</td><td>Voltage range, V</td><td>Exponent of a power</td><td>Voltage range, V</td><td>Exponent of a power</td></tr>
<tr><td rowspan="2">0,1</td><td>40-170</td><td>2.7±0.1</td><td>60-170</td><td>3.2±0.1</td></tr>
<tr><td>180-200</td><td>2.9±0.1</td><td>180-200</td><td>2.2±0.3</td></tr>
<tr><td rowspan="2">0,2</td><td>20-110</td><td>2.0±0.1</td><td>20-50</td><td>1.8±0.1</td></tr>
<tr><td>120-200</td><td>2.8±0.1</td><td>60-200</td><td>2.7±0.1</td></tr>
<tr><td rowspan="2">0,5</td><td>20-60</td><td>2.0±0.1</td><td>20-50</td><td>3.1±0.2</td></tr>
<tr><td>70-200</td><td>2.2±0.1</td><td>60-200</td><td>2.3±0.1</td></tr>
<tr><td rowspan="3">1</td><td>20-50</td><td>2.2±0.1</td><td rowspan="2">30-100</td><td rowspan="2">1.6±0.1</td></tr>
<tr><td>60-100</td><td>1.5±0.1</td></tr>
<tr><td>100-200</td><td>2.9±0.1</td><td>110-200</td><td>2.9±0.1</td></tr>
<tr><td rowspan="3">10</td><td>30-60</td><td>2.5±0.2</td><td>20-60</td><td>1.9±0.1</td></tr>
<tr><td>70-90</td><td>1.2±0.1</td><td>70-90</td><td>1.2±0.2</td></tr>
<tr><td>100-200</td><td>2.7±0.1</td><td>100-200</td><td>2.7±0.1</td></tr>
</table>

*S5. Simulations of the charge injection without domain formation*

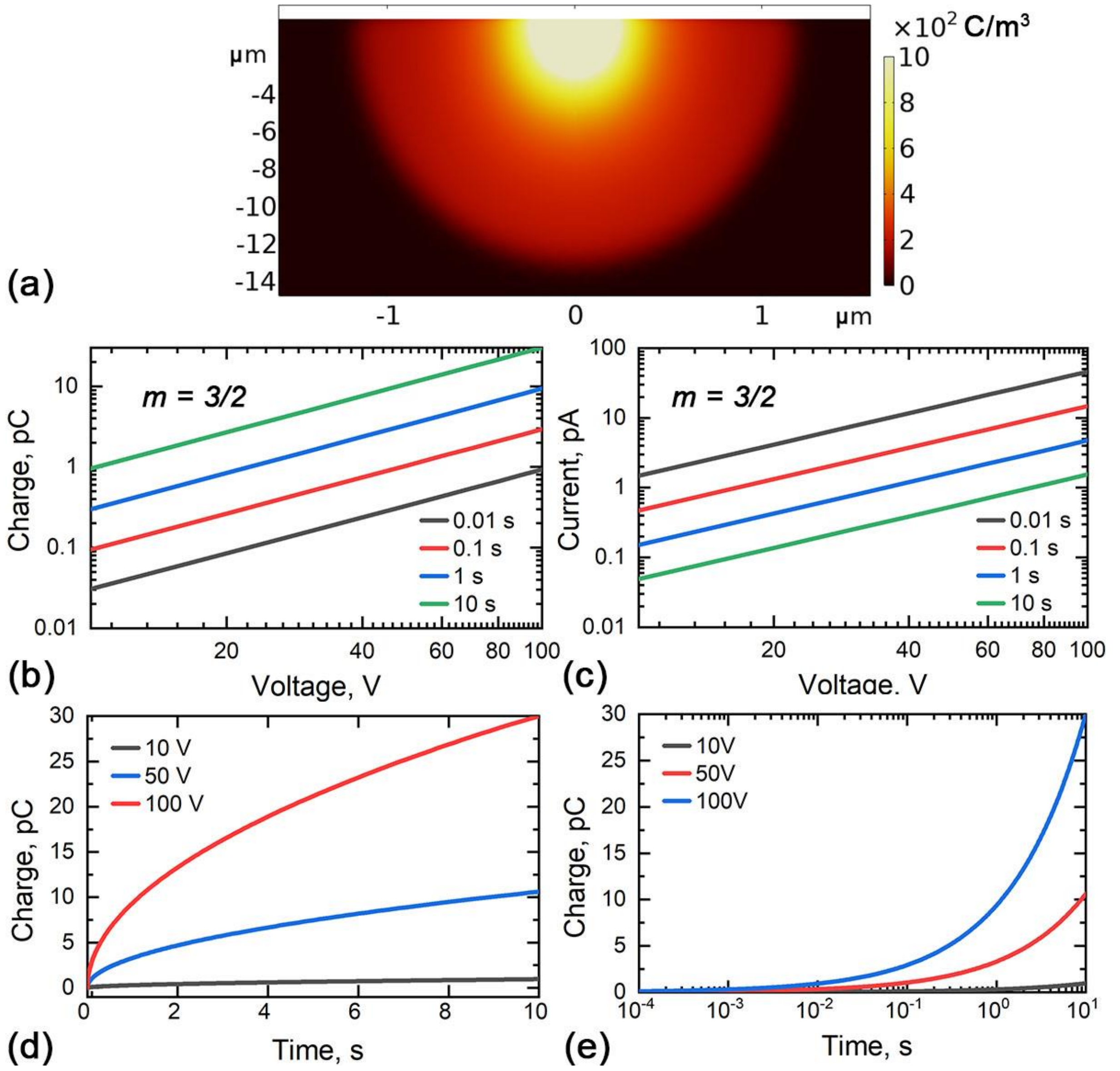

Figure S6. Simulation of charge injection into a dielectric under an applied bias. (a) Spatial map of the injected/accumulated charge in the near-contact region (brighter colors indicate higher charge density). (b) Total injected charge $Q$ versus voltage $V$ at different times follows power-law scaling $Q \propto V^m$ with $m \approx 3/2$. (c) Injection current $I$ versus voltage $V$at the same time instants, showing similar to charge behavior. (d) Time evolution of the injected charge for different applied voltages, approaching a quasi-steady value at long times. (e) Same charge–time dependence plotted on a logarithmic time axis to highlight the early-time transient regime and the voltage-dependent injection kinetics.

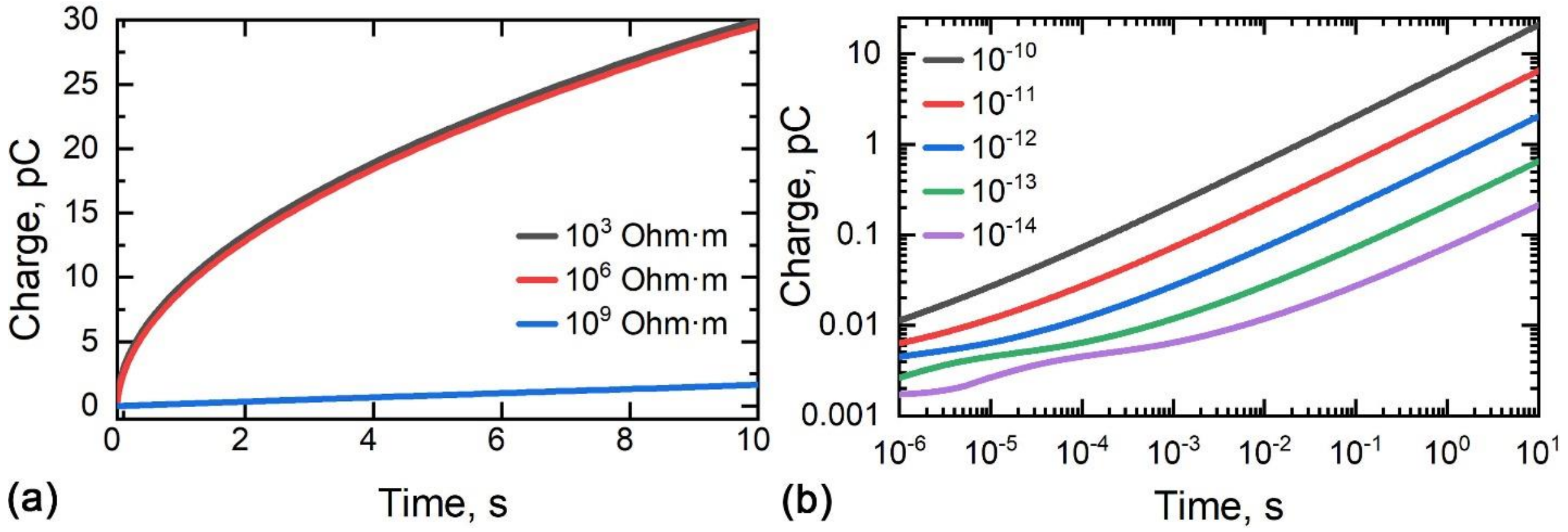

Figure S7. The dependence of the total injected charge on the (a) tip-sample contact resistance and on the (b) charge mobility.

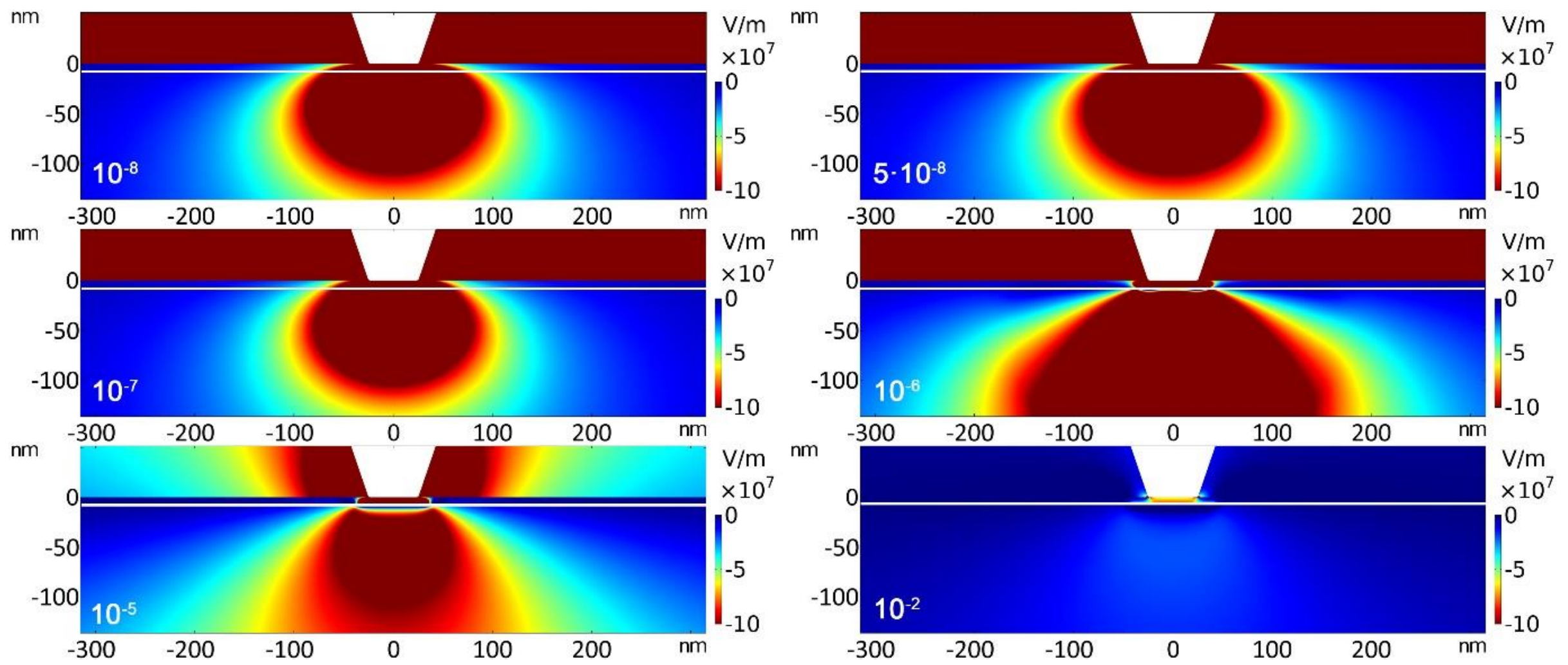

Figure S8. Spatial distribution of the electric field beneath the biased probe at the representative moments during the switching process. The snapshots illustrate how the field profile evolves as charge accumulates and redistributes in the material, as shown in Figure 6.

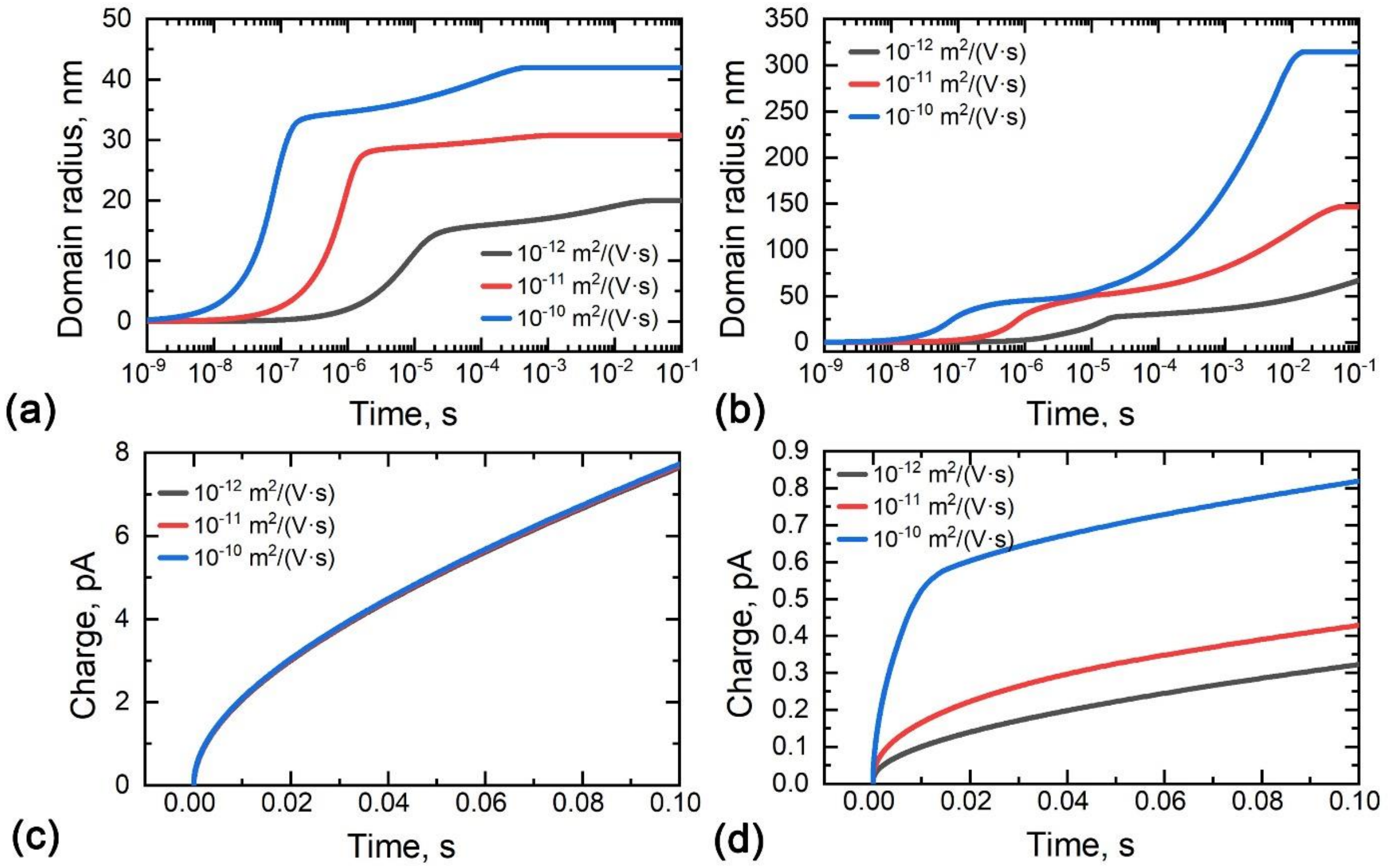

Figure S9. Evolution of the domain radius and accumulated charge for the different domain wall mobility, $\mu_{DW}$, indicated in the legend. Domain radius as a function of time for the different $\mu_e =$ (a) $10^{-10}$ and (b)$10^{-12}$ $m^2/(V{\cdot}s)$ bulk charge-carrier mobilities. (c,d) Corresponding temporal evolution of the accumulated injected charge.

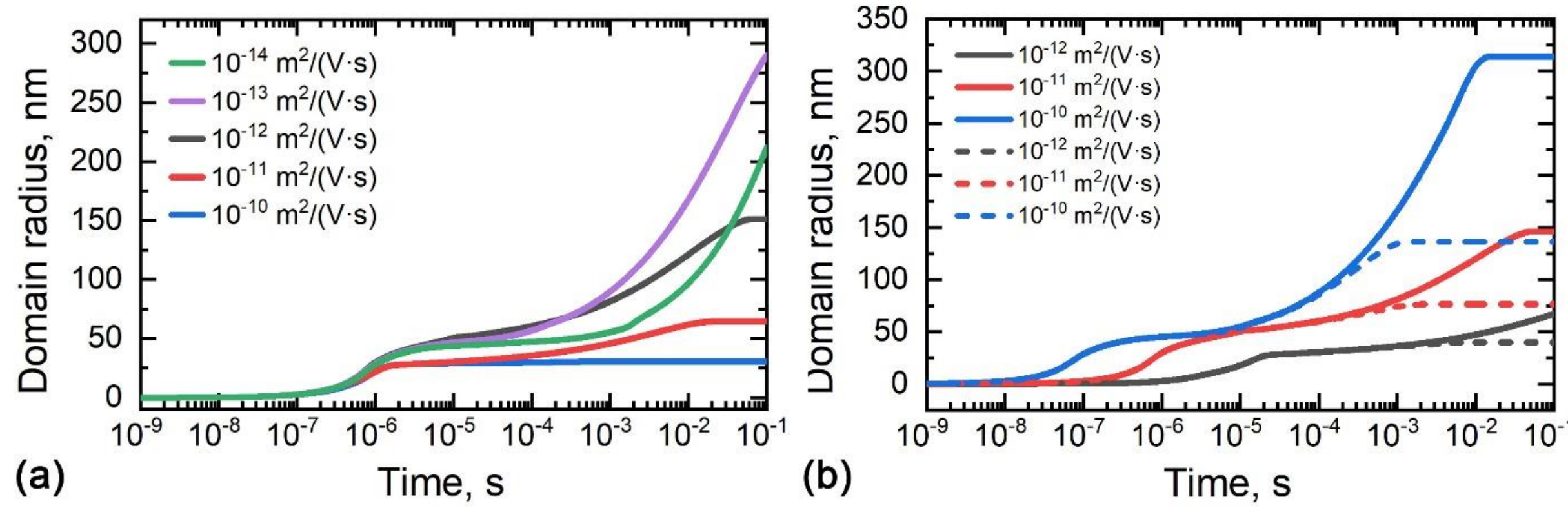

Figure S10. (a) Evolution of the domain radius as a function of time for different bulk charge-carrier mobilities, $\mu_e$, indicated in the legend, calculated for a fixed domain-wall mobility of $\mu_{DW} = 10^{-11}$ $m^2/(V{\cdot}s)$. (b) Evolution of the domain radius as a function of time for different domain-wall mobilities, $\mu_{DW}$, indicated in the legend. Solid and dashed curves correspond to threshold electric fields $E_{th}$ of 0.1 and 1 kV/mm, respectively.

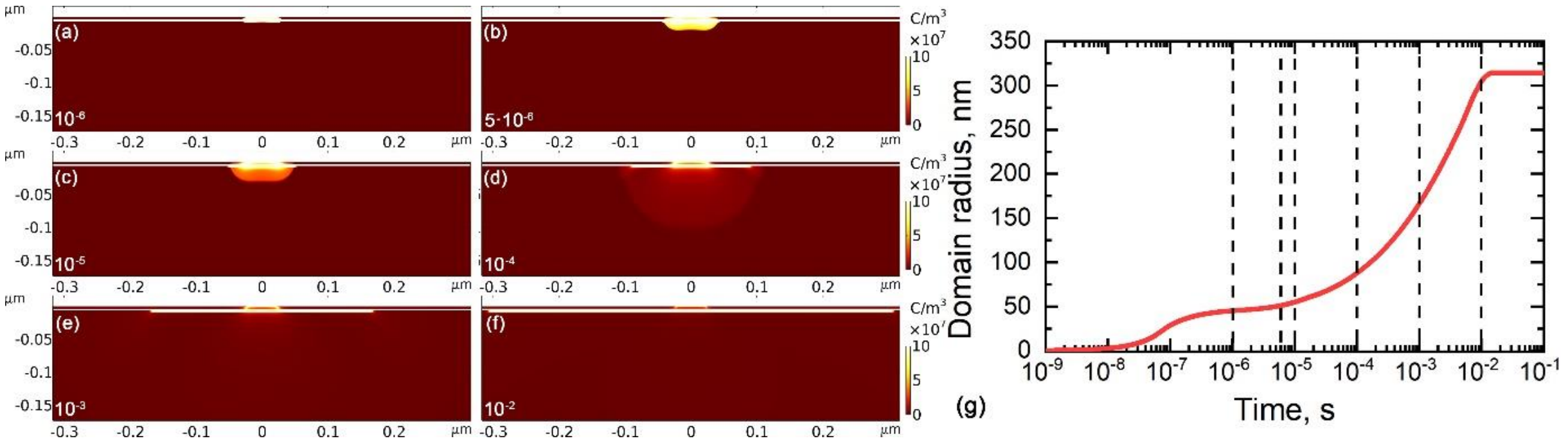

Figure S11. (a)-(f) Spatial distribution of the charged density beneath the biased probe at four representative moments of time during the polarization switching process and (g) respective temporal evolution of the domain radius. $\mu_e = 10^{-12}$ m$^2$/(V·s), $\mu_{DW} = 10^{-10}$ m$^2$/(V·s).

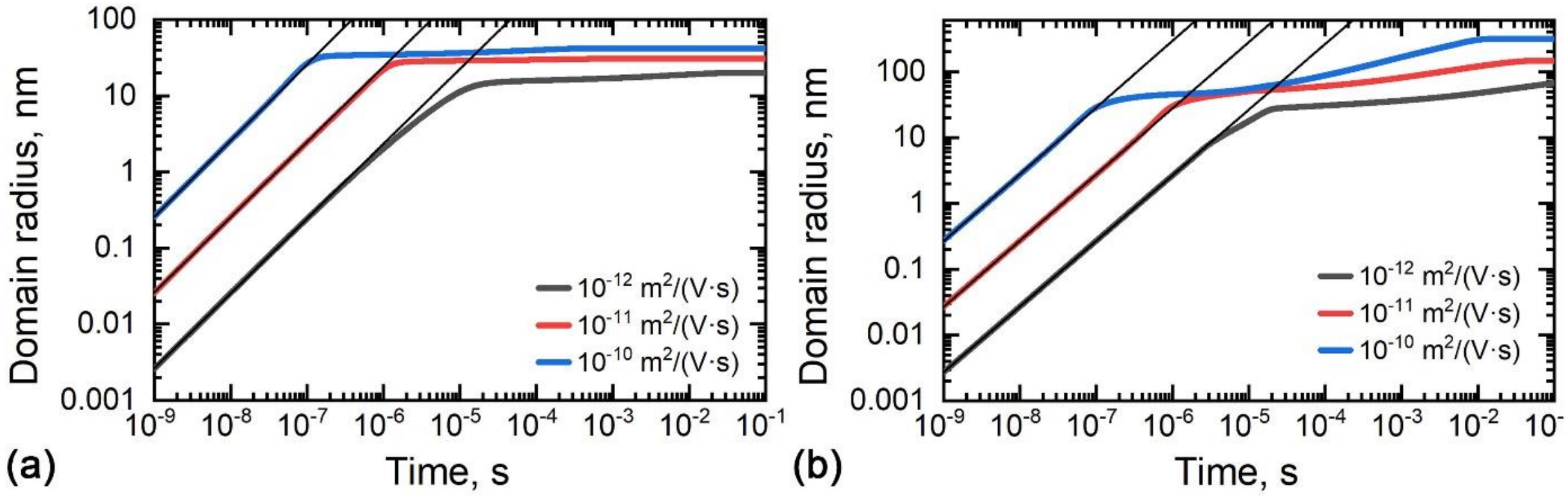

Figure S12. Evolution of the domain radius for the different domain wall mobility, $\mu_{DW}$, indicated in the legend. Domain radius as a function of time in a double-logarithmic scale for the different $\mu_e =$ (a) $10^{-10}$ and (b)$10^{-12}$ m$^2$/(V·s) bulk charge-carrier mobilities.